\input amstex
\input xy
\xyoption{all}
\input epsf
\documentstyle{amsppt}
\document
\magnification=1200
\NoBlackBoxes
\nologo
\hoffset=1.5cm
\voffset=2cm
\def\r{\roman}

\def\D{\Delta}

\def\A{\Cal{A}}

\def\Z{\bold{Z}_+}

\def\C{\bold{C}}
\def\H{\Cal{H}}

\pageheight{16cm}





 \centerline{\bf COMPLEXITY vs ENERGY:} 
 
\medskip

\centerline{\bf THEORY OF COMPUTATION AND THEORETICAL  PHYSICS \footnotemark1} 
\footnotetext{Talk at the satellite conference to ECM 2012, ``QQQ Algebra, Geometry, Information", Tallinn,
July 9--12, 2012.}

\bigskip

\centerline{\bf Yuri I. Manin}

\medskip

\centerline{\it Max--Planck--Institut f\"ur Mathematik, Bonn, Germany}

\bigskip

{\it ABSTRACT.}  This paper is a survey based upon the talk at the satellite 
conference to ECM 2012, ``QQQ Algebra, Geometry, Information", Tallinn,
July 9--12, 2012.  It is dedicated to the
analogy between the notions of {\it complexity} in theoretical computer science
and {\it energy} in physics. 
This analogy is not metaphorical: I describe three precise mathematical contexts,
suggested recently,
in which mathematics related to (un)computability is inspired by and to a degree reproduces
 formalisms of statistical physics and quantum field theory.

\bigskip

{\bf Contents}
\smallskip

 0. Introduction and summary

1. A brief guide to computability: operadic and categorical perspective

2.  Error--correcting codes and their  asymptotic bounds 

3. Zipf's law and Kolmogorov order

4. Feynman graphs and perturbation series in quantum physics

5. Graphs as flowcharts, and Hopf algebras

6. Regularization and renormalization 

\bigskip

\centerline{\bf 0. Introduction and summary}

\medskip

This  survey of several recent papers ([Man2], [Man4]--[Man7], [ManMar]) is dedi\-cated
to a deep analogy between the notions of {\it complexity} in theoretical computer science
and {\it energy} in physics. 

\smallskip
The analogy is not metaphorical: we describe several precise mathematical contexts,
suggested recently,
in which mathematics related to (un)computability is inspired by (and to a degree reproduces)
 formalisms of statistical physics and quantum field theory.

\smallskip

Namely, after recalling basics of the classical computability theory in sec. 1, we turn to three
main subjects:

\smallskip

a) {\it The problem of (un)computability of the asymptotic bound for error--correcting codes
over a fixed finite alphabet
(sec. 2).}
\smallskip
Here M.~Marcolli and the author have shown (in [ManMar] based upon [Man2])
 that the asymptotic bound arises as a phase transition curve
between different thermodynamic phases. The relevant
 partition function is a sum over the ensemble of all codes in which the role of energy
is played by the Kolmogorov complexity of the code.

\smallskip

b) {\it The problem of mathematical foundation for the empirical Zipf's law
(describing e.~g. the frequency distributions of words  in natural languages).}
\smallskip
It was suggested that this distribution reflects minimization of certain ``effort''.
I show that (in certain contexts) if this effort is defined as complexity,
Zipf's law emerges from L.~Levin's  {\it a priori} distributions,
mathematical theory of which was founded in the 1970's: see sec. 3
and more detailed argumentation in [Man7].

\smallskip

c) {\it The problem of uncomputable in the computability theory (sec. 4--6).} 

\smallskip
It is well--known that the theory of computability  unavoidably leads to effects
of uncomputability in its own realm: basically, it may be impossible
to decide in finite time whether a partial recursive function is defined at a given point.

\smallskip

I draw an analogy between this and problems of infinities in perturbative Quantum Field Theory.
Moreover, I suggest that renormalization schemes from QFT involving
first a deformation of the problem and then ``subtraction of infinities''
can be  fruitfully applied in computation theory. This procedure as well involves
Kolmogorov complexity. The basic common elements of the two
formalisms are graphs appearing as Feynman diagrams in QFT
and as flowcharts in computation theory. For more details
ad related results cf. [Man4], [Man5] and [Man3].
\smallskip

There is no proofs in this report: we focus on the presentation
of basic ideas.

\bigskip
\centerline{\bf 1. A brief guide to computability:}
\smallskip
\centerline{\bf  operadic and categorical perspective}

\medskip

Any single approach to mathematical notion of computability -- Turing's machines,
Church's lambda calculus,   Markov's algorithms -- 
by necessity bypasses rich intuitions governing other approaches. But since
this is unavoidable, and since our goal here is to pave the shortest way to Kolmogorov's
complexity, for us computability theory here will be based on the theory of (partial) recursive functions.
  
\medskip

{\bf 1.1. Three descriptions of partial recursive functions}.
A ``function'',  say,  $f:\,X\to Y$, below always means a pair   $(f, D(f))$, where
$D(f)\subset X$ and $f:\,D(f)\to Y$ a set--theoretic map. The definition domain is not always mentioned
explicitly. If $D(f)=X$, the function might be called ``total''; generally it may be called ``partial'' one.
The other extremal case is that of ``empty function'', with $D(f)=\emptyset$. We put $\bold{Z}_+:=\{1,2,3,\dots\}$.

\medskip

{\it (i) Intuitive description.} A function $f:\bold{Z}_+^m\to \bold{Z}_+^n$
is (partial) recursive iff it is ``semi--computable'' in the following sense:
 there exists an algorithm $F$ accepting as inputs vectors $x=(x_1,\dots ,x_m)\in \bold{Z}_+$
 with the following properties:
 
 \smallskip
 
 -- if $x\in D(f)$,  $F$ produces as output $f(x)$.
 
 \smallskip
 
 -- if $x\notin D(f)$, $F$ either produces the output ``NO'', or works indefinitely long without
producing any output.

\medskip

{\it (ii) Formal description (sketch).} It starts with two lists:

\smallskip

-- An explicit list of ``obviously'' semi--computable {\it basic
functions} such as constant functions,
projections onto $i$--th coordinate etc.

\smallskip

-- An explicit list of  {\it elementary operations},  performed over functions,
such as composition,  inductive definition, and  implicit definition by equation, that can be applied to several
semi--computable functions and ``obviously'' produces from them
a new semi--computable function.

\smallskip

After that, the set of {\it partial recursive functions}  is defined as the minimal set of functions $f:\bold{Z}_+^m\to \bold{Z}_+^n$,
with all $m,n\ge 0$,
containing all basic functions and closed wrt all elementary operations.  For details, see e.~g.~[Man1], Ch.~V.

\medskip
{\it (iii) Diophantine description (a difficult theorem).}   A function $f:\bold{Z}_+^m\to \bold{Z}_+^n$
is partial recursive iff there is a polynomial 
$$
P(x_1,\dots, x_m;y_1,\dots ,y_n; t_1,\dots ,t_q)\in \bold{Z}[x,y,t]
$$
such that the graph 
$$
\Gamma_f := \{(x,f(x))\} \subset  \bold{Z}_+^m\times \bold{Z}_+^n
$$
  is the projection  of the subset  $P=0$ in
$\bold{Z}_+^m\times \bold{Z}_+^n\times \bold{Z}_+^q$. For references and a proof, see e.~g~[Man1], Ch.~VI.

\medskip

{\bf 1.2. Constructive worlds.} An (infinite) {\it constructive world} is a countable set  $X$
(usually of some finite Bourbaki structures, such as the set of all error--correcting codes in
a fixed alphabet, cf. sec. 2.1 below) given together with a class of  
{\it structural\ numberings:} intuitively computable bijections
$\nu :\bold{Z}_+\to X$ which form a principal homogeneous space over the group
of totally recursive permutations of  $\bold{Z}_+$. A finite constructive world is any finite set.

\medskip
{\it Categorical Church's thesis, Part I.} Let $X$, $Y$ be two infinite constructive worlds, 
$\nu_X :\bold{Z}_+\to X$  $\nu_Y :\bold{Z}_+\to X$ their structural numberings, and  $F$ an (intuitive) algorithm 
that takes as input an object  $x\in X$ and produces an
object $F(x)\in Y$ whenever $x$ lies in the domain of definition of $F$; otherwise it outputs ``NO'' or works
indefinitely.

\smallskip

Then  $f:= \nu_Y^{-1}\circ F\circ\nu_X:\bold{Z}_+\to \bold{Z}_+$ is a partial recursive function.

\medskip

{\it Categorical  Church's thesis, Part II.} Let $\Cal{C}$ be a category, whose objects are
some infinite constructive worlds, and some finite constructive worlds
of all finite cardinalities. Define the set of morphisms $\Cal{C}(X,Y)$ to be the set
partial maps that can be algorithmically computed.

\smallskip

Then $\Cal{C}$ is equivalent to the category having one infinite object $\Z$, one finite object
$\{1,\dots ,\}$ of each cardinality, and partial recursive functions as morphisms.
If $X$ is finite, then  $\Cal{C}(X,Y)$ consists of all partial maps.

\medskip

{\bf 1.3. Kolmogorov complexity and Kolmogorov order.}
 Let $X$ be a constructive world. For any (semi)--computable function $u:\,\bold{Z}_+\to X$,
the (exponential)  complexity of an object $x\in X$ relative to $u$ is
$$
K_u(x):= \roman{min}\,\{m\in  \bold{Z}_+\,|\,u(m)=x\}.
$$
If such $m$ does not exist, we put   $K_u(x)= \infty .$

\smallskip
{\it Claim:} there exists such  $u$ (``an optimal Kolmogorov numbering'', or ``decompressor'') that
for each other $v:\,\bold{Z}_+\to X$, some constant $c_{u,v}>0$, and all $x\in X$,
$$
K_u(x)\le c_{u,v}K_v(x).
$$

 This $K_u(x)$ is called  {\it Kolmogorov complexity} of $x$.
\smallskip

A Kolmogorov order of a constructive world
$X$ is a bijection  $\bold{K}=\bold{K}_u:\, X\to \Z$ arranging elements of $X$
in the increasing order of their complexities $K_u$.

\medskip

Notice that any optimal numbering is only partial function, and its definition domain is not decidable.
Moreover, the Kolmogorov complexity $K_u$ itself is {\it not computable}: it is the lower bound
of a sequence of computable functions.

\smallskip

The same can be said about the Kolmogorov order.  Moreover, on $\bold{Z}_+$ it cardinally differs from the natural order
in the following sense:
it puts in the initial segments very large numbers that can be at the same time Kolmogorov simple.
For example, let $a_n:= n^{n^{.^{.^{.^{n}}}}}$ ($n$ times).  
Then $K_u(a_n)\le cn$ for some $c>0.$ 
\smallskip
In sec. 3 below we will discuss other remarkable properties of
complexity, in particular, its self--similar fractal properties.
\smallskip

Finally, the indeterminacy of the complexity related to different choices of optimal functions
$u,v$ is multiplicatively $\roman{exp}(\,O(1))$. The same is true for the Kolmogorov order.
\smallskip

For a thorough treatment of Kolmogorov complexity, cf. [LiVi]. Notice that in the literature
one often uses the {\it logarithmic} Kolmogorov complexity which is defined
as the length of the binary presentation of $K_u(x)$. It is interpreted as the length of
the {\it maximally compressed description} of $x$. For our purposes, exponential
version is more convenient, in particular, because it allows us to define an unambiguous
Kolmogorov order on $\Z$ or any infinite constructive world.

\medskip

{\bf 1.4. Oracle assisted computations.} The formal description of partial recursive functions
in sec. 1.1 (ii) allows one to define larger classes of partial functions that
can be obtained by {\it oracle assisted} computations. The point is that the standard
elementary operations can be applied to {\it arbitrary} partial functions. Therefore
we can add any uncomputable (not partial recursive) functions to the list
of basic functions and consider the minimal subset of partial functions
containing this expanded list and closed wrt elementary operations.
\smallskip

This option was used in [Man7] in order to define the respective extensions of the notion 
of complexity and apply them to the explanation of Zipf's law in the situations,
related to oracle assisted computations and library reuse, cf. sec.~3 below.
\smallskip

Formally, we are considering the (pro)perad generated by the elementary operations on partial functions
and various algebras over it. It would be important to understand all relations
between elementary operations. For the first steps in this direction,
cf. [Ya]; for a general formalism, cf. [BoMan].

\vskip1cm

\centerline{\bf  2.  Error--correcting codes and their  asymptotic bounds }

\medskip

{\bf 2.1. Basic notation.} Choose an alphabet $A$, a finite set of cardinality $q\ge 2$.
A {\it code}  $C\subset A^n$ is a subset of words of length $n$.
{\it Hamming distance} between two words of the same length is defined as
$$
d((a_i),(b_i)):= \roman{card} \{i\in (1,\dots ,n)\,|\,a_i\ne b_i\}.
$$
{\it Code parameters} are the  cardinality of the alphabet  $q$ and the numbers $n(C), k(C), d(C)$ defined by:
$$
n(C):=n,\quad  k(C):=k:= [\roman{log}_q \roman{card}(C)],\quad 
$$
$$
d(C) := d= \roman{min}\,\{d(a,b)\,|\,a,b\in C, a\ne b\}.
$$
Briefly,  $C$ is an  $[n,k,d]_q$--code. Its {\it code point} is the point
$$
x(C):= \left( \frac{k(C)}{n(C)}, \frac{d(C)}{n(C)},\right)\in [0,1]^2
$$
Coordinates of $x(C)=(R(C),\delta (C))$ are called  {\it transmission rate} and  {\it relative distance} respectively.

\smallskip
The idealized scheme of using error--correcting codes for information transmission 
can be described as follows. Some source data are encoded by a sequence of code words.
After transmission through a noisy channel at the receiving end we will get a sequence of possibly corrupted words.
If we know probability of corruption of a single letter, we can calculate, how many corrupted letters in a word
we may allow for safe transmission; pairs of code words  must be then separated by
a larger Hamming distance. This necessity puts an upper bound on the achievable transmission rate. 
\smallskip
{\it A good code}  must maximize minimal relative distance when the transmission rate is chosen.

\smallskip

Our discussion up to now was restricted to {\it unstructured} codes: arbitrary subsets of words.
Arguably, one more property of good codes is the existence of efficient algorithms of encoding
and decoding. This can be achieved by introduction of {\it structured} codes. A typical choice
is represented by {\it linear codes}: for them, $A$ is a finite field of $q$ elements, and $C$ is a linear subspace
of   $\bold{F}_q^n$.
\medskip

{\bf 2.2. Asymptotic bound.} Call  {\it the multiplicity} of a code point the number of codes that project onto it.

\medskip

{\bf 2.2.1. Theorem.} (Yu.~M., 1981 + 2011). {\it  There exists a continuous function $\alpha_q (\delta)$, $\delta\in[0,1]$,
with the following properties: 

\smallskip

(i) The set of code points of infinite multiplicity is exactly the set of  rational points  $(R,\delta )\in [0,1]^2$
satisfying $R\le \alpha_q(\delta )$. 
\smallskip
The curve  $R= \alpha_q(\delta )$ is called the asymptotic bound.
\smallskip

(ii) Code points $x$ of finite multiplicity all lie strictly above  the asymptotic bound and are
called isolated\ ones: for each such point  there is an open neighborhood  containing $x$ as the only code point.
\smallskip

(iii) The  same statements are true for linear codes, with a possibly different
asymptotic bound $R= \alpha_q^{lin} (\delta )$.}

\medskip

{\bf 2.3. Can one compute an asymptotic bound?} During the thirty years since the discovery of the asymptotic bounds,
many upper and lower estimates were established for  them, especially for the linear case: see the monograph [VlaNoTsfa].
Upper bounds helped to pinpoint a number of isolated codes.

\smallskip

However, the following most natural problems remain unsolved:
\smallskip
-- To find an explicit formula for  $\alpha_q$ or $\alpha_q^{lin}$.

\smallskip

-- To find any single value  of $\alpha_q(\delta )$ or $\alpha_q^{lin}(\delta )$ for $0<\delta <1-q^{-1}$
(at the end segment $[1-q^{-1},1]$ these function vanish).

\smallskip

-- To find any method of approximate computation of   $\alpha_q(\delta )$ or $\alpha_q^{lin}(\delta )$.

\medskip
-- Clearly, $\alpha_q^{lin} \le  \alpha_q$. Is this inequaliy {\it strict} somewhere?

\medskip
{\bf 2.4. A brief survey of some known results.} (i) One can count the number of codes of bounded
block length $n$ and plot their code points. The standard probabilistic methods then
give the following {\it Gilbert--Varshamov bounds}.

\smallskip
Most unstructured $q$--ary codes lie lower or only slightly above the Hamming
curve
$$
R=1-H_q(\delta /2),
$$
$$
H_q(\delta ) =\delta\roman{log}_q (q-1)-\delta \roman{log}_q \delta -
(1-\delta ) \roman{log}_q (1-\delta ).
$$
 Most linear $q$--ary codes lie near or only slightly above the Gilbert--Varshamov bound 
$$
R=1-H_q(\delta).
$$ 
In particular,
$$
\alpha_q (R)\ge \alpha_q^{lin}(R)\ge 1-H_q(\delta)
$$

(ii) A useful combinatorial upper estimate is the Singleton bound:
$$
R(C)+\delta (C) \le 1+ \frac{1}{n(C)}.
$$
Hence
$$
\alpha_q (\delta)\le 1-\delta .
$$
It follows that code points lying above this bound are isolated. The following
Reed--Solomon (linear) codes $C\subset  \bold{F}_q^n$ belong to this group.
\smallskip

Choose parameters   $1\le k\le n\le q, d=n+1-k$.
Choose pairwise distinct $x_1,\dots ,x_n\in \bold{F}_q$,
Embed the space of polynomials $f(x)\in \bold{F}_q[x]$ of degree $\le k-1$ into
$\bold{F}_q^n$ by
$$
f\mapsto (f(x_1),\dots , f(x_n))  \in \bold{F}_q^n .
$$

\smallskip

After works of Goppa, this construction was generalized. Points $x_1,\dots ,x_n\in \bold{F}_q$
were replaced by rational points of any smooth algebraic curve over  $\bold{F}_q$,
and polynomials by sections of an invertible sheaf. This allowed one to construct non--isolated linear
codes lying partly strictly above the Gilbert--Varshamov bound.
\smallskip

This implies that  we cannot  ``see'' the asymptotic bound, plotting  the set of (linear)  code points of bounded size:
we will see a cloud of points, whose upper bound concentrates near the Hamming or Varshamov--Gilbert bounds.

\medskip

{\bf 2.5. Partition function for codes involving complexity.} The situation drastically changes,
at least theoretically, if we allow ourselves to rearrange the codes in the order
of growing Kolmogorov complexity.

\smallskip

In order to state our principal theorem, notice that the function $\alpha_q(\delta )$ is continuous
and strictly decreasing for $\delta \in [1,1-q^{-1})$. 
Hence the limit points domain $R\le  \alpha_q(\delta )$
can be equally well described by  the inequality $\delta\le \beta_q(R)$
where $\beta_q$ is the function inverse to $\alpha_q$.

\smallskip

Fix  an $R\in \bold{Q} \cap (0,1)$. For $\Delta \in  \bold{Q} \cap (0,1)$, put
$$
Z(R, \Delta;\beta ):=   \sum_{C:\, R(C)=R,\,\Delta\le \delta (C) \le 1} K_u(C)^{-\beta +\delta (C) -1},
$$
where $K_u$ is an (exponential) Kolmogorov complexity on the constructive world of all codes
in a given alphabet of cardinality $q$.

\bigskip

{\bf 2.6. Theorem.}  {\it (i) If $\Delta > \beta_q(R)$, then $Z(R, \Delta;\beta )$ is a real analytic function of $\beta$.

\smallskip

(ii) If $\Delta < \beta_q(R)$,  then $Z(R, \Delta;\beta )$ is a real analytic function of $\beta$ 
for $\beta >\beta_q(R)$ such that its limit for $\beta -\beta_q(R)\to +0$ does not exist.}

\medskip

The following thermodynamical analogies justify our interpretation of the asym\-ptotic bound 
a phase transition curve.
\smallskip
a) The argument $\beta$ of the partition function corresponds to the inverse tempe\-rature.

\smallskip

b) The transmission rate $R$ corresponds to the density $\rho$.

\smallskip

c) Our asymptotic bound
transported into $(T=\beta^{-1},R)$--plane as  $T=\beta_q(R)^{-1}$ becomes
the phase transition boundary in the (temperature, density)--plane.

\bigskip

\centerline{\bf 3. Zipf's law and Kolmogorov order}

\medskip

{\bf 3.1.~Zipf's law.} G.~Zipf studied the frequencies with which words of a natural language
are used in various texts. He found a remarkably stable pattern ([Zi1], [Zi2]):
if all words $w_k$  of a language
are ranked according to decreasing frequency of their appearance 
in a representative corpus of texts, then the frequency $p_k$ of $w_k$ is approximately inversely proportional
to  its rank $k$: see e.~g.~ Fig.~1 in [Ma1] based upon a corpus containing $4\cdot 10^7$ Russian words.

\smallskip
Zipf himself has suggested that this distribution ``minimizes effort''. Mandelbrot 
in [Mand] has shown that if we postulate and denote by $C_k$ a certain ``cost'' (of producing, using etc.) of the
word of rank $k$, then the frequency distribution $p_k \sim 2^{-h^{-1}C_k}$ minimizes
the ratio $h=C/H$, where $C:=\sum_k p_kC_k$ is the average cost per word,  and $H:= -\sum_kp_k \roman{log}_2 p_k$
is the average entropy: see [Ma2].

\smallskip

We get from this a power law, if  $C_k\sim \roman{log}\,k$.  An additional problem,
what is so special about power $-1$, must be addressed separately. 
\smallskip

In all such discussions, it is more or less implicitly assumed that empirically observed distributions
 concern  fragments of a potential countable infinity of objects.  In the mathematical model
 suggested in [Man7] it is assumed  that these objects
 form an infinite constructive world in the sense of 1.2 above. Below I will survey this model.

\medskip

{\bf 3.2.~How minimization of complexity leads to Zipf's law.} A mathematical model of Zipf's law is based upon two
postulates:

\smallskip

{\it (A) Rank ordering coincides with a Kolmogorov  ordering  
(up to a factor  $exp\,(O(1))$), cf. 1.3 above.}

\smallskip

{\it (B) The probability distribution producing Zipf's law (with exponent $-1$) is
(an approximation to) the L.~Levin maximal computable from below
distribution: see [ZvLe], [Lev1], [Lev2] and [LiVi].}

\medskip
If we accept $(A)$ and $(B)$, then Zipf's law follows from two basic properties of Kolmogorov complexity:
\medskip
{\it (a) rank of $w$ defined according to (A) is $exp\,(O(1))\cdot K(w)$.}
\medskip

{\it (b) Levin's distribution  assigns to an object $w$  probability
$\sim KP(w)^{-1}$ where $KP$ is the exponentiated prefix Kolmogorov complexity (cf. [LiVi], [CaSt]),
and we have, up to  $exp\, (O(1))$--factors,
$$
K(w)\preceq KP(w)\preceq  K(w)\cdot \roman{log}^{1+\varepsilon} \,K(w)$$}
with arbitrary $\varepsilon >0$.
\smallskip

There is a slight discrepancy between the growth orders of  $K$ and $KP$. This discrepancy ensures
the convergence of the series $\sum_w KP(w)^{-1}$. On finite sets of data 
this small discrepancy is additionally masked by the dependence
of both $K$ and $KP$ on the choice of  an optimal encoding. 
\smallskip

``Minimization of effort'' is thus achieved if  effort itself is interpreted as 
the  length of the maximally
compressed prefix  free description of  an object. 
\smallskip

Such a picture makes sense especially if the objects satisfying
Zipf's distribution, are {\it generated} rather than simply {\it observed.}

\smallskip
This matches very well the results of the previous section on asymptotic bounds
for error--correcting codes: if one produces codes in the order of their Kolmogorov
complexity rather than size, their code points will well approximate the picture of the whole domain
under the asymptotic bound. Moreover, Levin's distribution very naturally leads to
the thermodynamic partition function on the set of codes, and to the interpretation of asymptotic bound as
a phase transition curve.  In sec. 2, we have written it in the form the form $\sum_C K(C)^{-s(C)}$ where $s(C)$
is a certain function defined on codes and including as parameters analogs of temperature and density. 
We could replace $K$ with $KP$, and freely choose
the optimal family defining complexity: this would have no influence at all on the form of the phase curve/asymptotic bound.
\smallskip

It is interesting to observe that the mathematical problem of generating good error--correcting codes
historically made a great progress in the 1980's with the discovery of algebraic geometric
Goppa codes, that is precisely with the discovery of greatly compressed descriptions of
large combinatorial objects.

\smallskip

To summarize, the class of a priori probability distributions that we are considering here
is {\it qualitatively distinct} from those that form now  a common stock of sociological and sometimes scientific analysis:
cf.~a beautiful synopsis of the latter by Terence Tao in [Ta] who also stresses that
``mathematicians do not have a fully satisfactory and convincing explanation
for how the [Zipf] law comes about and why it is universal''.

\smallskip

What arguments could furnish such an explanation? Ubiquity of Gaussian distribution, for example,
is often explained away by appealing to the central limit theorem: average of many independent
random (equally distributed) variables  tends to be Gaussian for whatever initial distribution. 
Below I will  argue that  universality of Zipf's law is similarly based on the surprisingly
self--similar nature of Kolmogorov complexity.
\medskip

{\bf 3.3. Fractal landscape and self--similarity of the Kolmogorov complexity.} 
In [LiVi], pp. 103, 105, 178, one can find a schematic graph of logarithmic complexity of naturals.
The visible``continuity" of this graph
reflects the fact that complexity of $k+1$ in any reasonable encoding is almost the same
as complexity of $k$. It looks as
follows: most of the time it follows closely  the graph of $\roman{log}\,k$,
but infinitely often it drops down, lower than any given computable function:

\smallskip

One does not see or suspect
self--similarity. But it is there:
if one restricts this graph onto any infinite decidable subset
of $\Z$ in increasing order,
one will get the same complexity relief as for the whole $\Z$:
in fact, for any recursive bijection $f$ of $\Z$ with a subset of $\Z$
we have $K(f(x)) = exp (O(1))\cdot K(x)$.

\smallskip

If we pass from complexity to a Levin's distribution, that is, basically, invert the 
values of complexity, these fractal properties survive. 

\smallskip

This property can be read as the extreme stability of such a distribution
with respect to the passage to  various sub--universes
of objects, computable renumbering of objects etc.,
in the same way as the picture of random noise
in a stable background is held responsible for universality
of normal distribution.

\medskip

{\bf 3.4.  Complexity  on the background of oracle assisted computations and library reuse.}
In the paper [Ve],   T.~Veldhuizen  considers Zipf's law in an unusual context that did not exist in the days 
when Kolmogorov, Solomonov and Chaitin made their ground--breaking discoveries, but which
provides, in a sense, landscape for an industrial incarnation of  complexity.
Namely, Veldhuizen studies actual software and software libraries and 
analyzes  possible profits from software reuse. Metaphorically, this is a picture of human culture
whose everyday existence depends on a continuous reuse of treasures
created by researchers, poets, philosophers, cf. [Man6].
\smallskip

Mathematically, reuse furnishes new tools of
compression:  roughly speaking, a function $f$ may have a very large
Kolmogorov complexity, but the length of the library address of its program
may be short, and only the latter counts 
if one can simply copy the program from the library. 

\smallskip

In order to create a mathematical model of reuse and its Zipf's landscape, 
the notion of an admissible set of partial functions 
note, I need to define the mathematical notion
of {\it relative Kolmogorov complexity $K(f|\Phi)$.} 

\medskip

{\bf 3.5. Admissible sets of functions.} Consider a  set $\Phi$  of partial functions   $f:\, (\Z^{+})^m \to  (\Z^{+})^n$, $m,n\ge 0$.
We will call $\Phi$ {\it an admissible set}, if it is countable and satisfies the following conditions. 

\smallskip
{\it (i) $\Phi$   is closed under composition and contains all projections (forget some coordinates),
and embeddings (permute and/or add some constant coordinates).}

\smallskip

 Any $(m+1,n)$--function  can be considered as a family of $(m,n)$--functions $(u_k)$:
$u_k(x_1,\dots ,x_m):= u(x_1,\dots ,x_m, k)$.  From (i) it follows that for any $u\in \Phi$ and $k\in \Z^+$, also
$u_k\in \Phi$. Similarly, if $u(x_1,\dots ,x_m)$ is in $\Phi$, then 
$$
U(x_1,\dots,x_m,x_{m+1},\dots ,x_{m+n})\equiv
u(x_1,\dots ,x_m)
$$ 
is in $\Phi$.

\smallskip
{\it (ii) For any $(m,n)$, there exists 
such an $(m+1,n)$--function $u\in \Phi$  that the  family of functions   $u_k:\,(\Z^{+})^m \to  (\Z^{+})^n$, 
contains all $(m,n)$--functions   belonging to $\Phi$.
\smallskip
We will say that such a function $u$ (or family $(u_k)$) is ample.}

\smallskip

{\it (iii) Let $f$ be a total recursive function $f$ whose image is
 decidable, and $f$ defines a bijection between $D(f)$ and  image of $f$.
 Then $\Phi$ contains both $f$ and $f^{-1}$.}

\smallskip

It is shown in  [Man7] that one can define analog of complexity with respect to such a set,
$K(x\,|\,\Phi )$ and, moreover, that such sets can be obtained as ``algebras'' over
a (pro)perad generated by standard operations that usually are applied only to
partially recursive functions.

\smallskip

There are many instances of empiric Zipf's laws where our picture might
be applicable: cf. [Del], [DeMe], [De], [MurSo].
 Such a reduction of the Zipf law for  natural languages
might require for its justification some  neurobiological data: cf. [Ma1], appendix A
in the arXiv version.

\bigskip

\centerline{\bf 4. Feynman graphs and perturbation series in quantum physics}
\medskip

{\bf 4.1. A toy model.} {\it Feynman path integral} is an heuristic expression of the form
$$
\frac{\int_{\Cal{P}}e^{S(\varphi )}D(\varphi )}{\int_{\Cal{P}}e^{S_0(\varphi )}D(\varphi )}
\eqno(4.1)
$$
or, more generally,  a similar heuristic expression for {\it correlation functions}.
\smallskip
In the expression (4.1), $\Cal{P}$ is imagined as a functional space
of {\it classical fields $\varphi$} on  a {\it space--time manifold} $M$; 
$S:\,\Cal{P}\to \bold{C}$ is a functional of {\it classical action} measured in Planck's units.
$S_0$ is its {\it quadratic part}, or {\it ``free field action''.}
\smallskip

Usually $S(\varphi )$ itself is
an integral over $M$ of a local density on $M$ called {\it Lagrangian.}
In our notation  
$S(\varphi )=-\int_M L(\varphi (x)) dx.$ Lagrangian density may depend on derivatives, include
distributions etc.  

\smallskip
Finally, the integration measure $D(\varphi )$ and the integral itself $\int_{\Cal{P}}$
should be considered as symbolic constituents of the total expression (4.1) conveying
a vague but powerful idea of  {\it  ``summing quantum amplitudes over virtual
classical trajectories''. }

\smallskip

In our toy model, we will replace $\Cal{P}$ by a finite--dimensional
real space. We endow it with a basis indexed by a finite set
of ``colors''  $A$, and an Euclidean metric $g$ encoded by the symmetric
tensor $(g^{ab}),\,a,b\in A.$ We put $(g^{ab})=(g_{ab})^{-1}.$

\smallskip

The action functional $S(\varphi )$ is a formal series in linear coordinates on $\Cal{P}$, $(\varphi ^a)$,
of the form
$$
S(\varphi )=S_0 (\varphi) + S_1(\varphi ),\quad
S_0(\varphi ):=-\frac{1}{2} \sum_{a,b} g_{ab}\varphi^a\varphi^b,
$$
$$
S_1(\varphi ):=\sum_{k=1}^{\infty}\frac{1}{k!}\sum_{a_1,\dots ,a_k\in A}
C_{a_1,\dots ,a_k}\varphi^{a_1}\dots \varphi^{a_k}
\eqno(4.2)
$$
where $(C_{a_1,\dots ,a_n})$ are certain  symmetric tensors.

\smallskip
 Below we will consider $(g_{ab})$ and  
$(C_{a_1,\dots ,a_n})$ as independent formal variables, 
{\it ``formal coordinates on the space of theories''.}

\smallskip 
We will express the toy version of (4.1)
as a formal series over (isomorphism classes of)  graphs.

\smallskip
A (combinatorial) graph $\tau$, by definition,  consists of two finite sets: flags $F_{\tau}$
and vertices $V_{\tau}$.  Besides, an involution $j_{\tau}$ of $F_{\tau}$ is given, showing which
pairs of flags form halves of edges, and which are not (tails). Finally, the map $\partial_{\tau}:\,F_{\tau}\to V_{\tau}$
shows to which vertex each graph is incident.
The geometric realization of $\tau$ is a topological space whose structure is suggested by
the choice of words in the definition:

  \medskip

$$
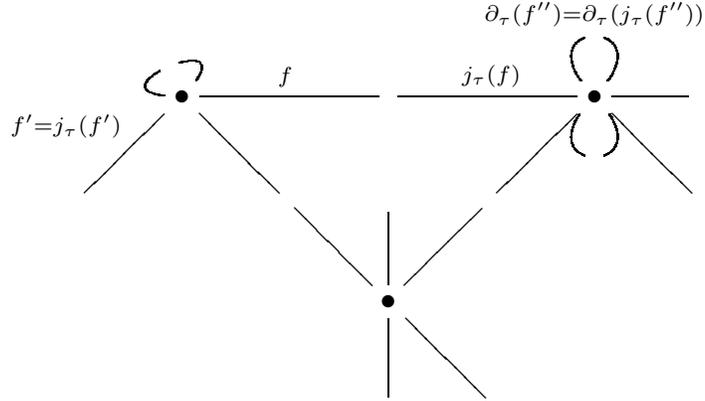
\xymatrix{
& \bullet\ar@{-}[rr]^f
\ar@{-}[dr]\ar@{-}[dl]_{f'=j_{\tau}(f')}\ar@{-}@(l,ur)[]|\hole&& \ar@{-}[rr]{ }\ar@{-} [rr]^{j_{\tau}(f)} && \bullet\ar@{-}[r]\ar@{-}[dr] \ar@{-}[dl]
\ar@{-}@(ur,ul)[]|\hole_{\partial_\tau (f'')=\partial_\tau (j_\tau (f''))} \ar@{-}@(dr,dl)[]|\hole&
{ }
\\
{ }&& { } &{ }& { } && { }
\\
&&&\bullet\ar@{-}[d]\ar@{-}[rd]\ar@{-}[ur]\ar@{-}[ul]\ar@{-}[u]\\
&&& { }&{ }
}
$$

Each edge $e$ consists of a pair of flags denoted
$\partial{e}$, and each vertex $v$ determines the set  of flags
incident to it denoted $F_{\tau} (v)$. By  $\chi (\tau )$ we denote the 
Euler characteristic of the geometric realization of $\tau$.

\bigskip

{\bf 4.2. Theorem.} {\it  Let  $\lambda$ be a formal parameter. Then
$$
\frac{\int_{\Cal{P}}e^{\lambda^{-1}S(\varphi )}D(\varphi )}{\int_{\Cal{P}}
e^{\lambda^{-1}S_0(\varphi )}D(\varphi )} =
  \sum_{\tau\in\Gamma}\frac{\lambda^{-\chi (\tau )}}{|\roman{Aut}\,\tau |}\,
w(\tau )
\eqno(4.3)
$$
where $\tau$ runs over isomorphism classes 
of all finite graphs $\tau$. The weight $w(\tau )$ of such a graph
is determined by the action functional (1.2) as follows:
$$
w(\tau ):=\sum_{u:\,F_{\tau}\to A}\ \prod_{e\in E_{\tau}}
g^{u(\partial e)}\prod_{v\in V_{\tau}} C_{u(F_{\tau}(v))}\,.
\eqno(4.4)
$$
}
\bigskip
More precisely, the identity (4.2) is  obtained by first  interpreting the 
integrands in the numerator of (4.2) as formal series in $(g_{ab}, C_{a_1,\dots,a_k})$, and
then integrating term--wise by using the well known formulas
for Gaussian integrals. 

\bigskip

\centerline{\bf 5. Graphs as flowcharts, and Hopf algebras}
\medskip

{\bf 5.1. Graphs as flowcharts.}  Feynman diagrams of more realistic models
 and graphs  used in the computation theory can be considered as {\it flowcharts}
 describing the flow of information from a part of tails playing role of
 {\it inputs} to another part, playing role of {\it outputs}. At vertices,
 the information gets processed.
 
 \smallskip
 
 In order to make such an interpretation workable, we need pay more attention
 to orientation. 
{\it Orientation} of a graph $\tau$ is the decoration $F_{\tau} \to L_F= \{in, out\}$ such that
halves of any edge are decorated by different labels.

\smallskip

Tails of $\tau$ oriented $in$ (resp. $out$) are called
{\it (global) inputs $T_{\tau}^{in}$} (resp. {\it (global) outputs} $T_{\tau}^{out}$) of $\tau$.
Similarly, $F_{\tau}(v)$ is partitioned into inputs and outputs
of the vertex $v$.

\smallskip

An oriented graph $\tau$ is called {\it directed} if it
satisfies the following condition:
\smallskip
 On each connected component 
one can define a continuous real valued function (``time'')
in such a way that
moving in the direction of orientation along each flag
inreases the value of this function.
\smallskip
In particular, oriented trees and forests are always directed,
and physical Feynman diagrams without loops  as well. 

\smallskip

 An abstract {\it flowchart} is a directed graph endowed with the decoration
of its vertices by a set $Op$ of (names of) operations
that can be performed on certain inputs producing certain outputs.
Generally, flags are also labeled by {\it types} of the arguments.

\smallskip

To be more precise, flowcharts in theoretical computer science form a natural hierarchy.
\smallskip
At the lower level of this hierarchy, {\it histories of computations} are situated.
For example, the sequence of the states of a Turing machine, performing
a concrete computation, may be encoded by a flowchart, in which
inputs of all vertices are decorated by $0$ or $1$, and vertices themselves
carry either  the name of identical operation or
the name of the internal state of the head, reading the respective site.
Such a history may well be {\it infinite}.

\smallskip
At  higher  levels flowcharts may serve
as {\it descriptions}:  programs represented as compositions of some 
subprograms, but not specifying concrete values of arguments and thus hiding
the actual computation process and/or compressing the notation.

\smallskip

We omit here a formal definition of admissible sets of decorated flowcharts:
cf. [Man4], [Man5] for further details. Briefly, an admissible set must  be closed wrt finite
disjoint unions and {\it cuts} that will be defined below.

\smallskip

For another version of flowcharts, see [Sc].

\medskip

{\bf 5.2. Connes--Kreimer bialgebras of flowcharts} ([ConKr]).  Let
$Fl$ be an admissible set of decorated graphs,  $k$ := a commutative ring.
We denote by $H=H_{Fl}$ the $k$--linear span of isomorphism classes $[\tau ]$ of graphs $\tau$ in $Fl$
and define multiplication by
$$
m :\, H\otimes H\to H,\quad m ([\sigma ]\otimes [\tau ]):= [\sigma \coprod \tau ],\quad
$$
We pass now to cuts and comultiplication.

\smallskip

Let $\tau$  be an oriented graph. Call {\it a proper cut} $C$
of $\tau$ any partition of $V_{\tau}$ into a disjoint union of two non--empty subsets 
$V_{\tau}^C$ (upper vertices) and $V_{\tau,C}$ (lower vertices) satisfying the following conditions:  

\smallskip
(i) For  each oriented wheel in $\tau$, all its vertices belong either to $V_{\tau}^C$,
or to $V_{\tau, C}$.

\smallskip

(ii) If an edge $e$ connects a vertex $v_1\in V_{\tau}^C$ to 
$v_2 \in V_{\tau,C}$, then it is oriented from $v_1$ to $v_2$
(``information flows only from past to future'' ).

\smallskip

(iii) Two improper cuts: $\tau^C:=\tau$ or $\tau_C=\tau$.

\smallskip

Denote by $\tau^C$  (resp. $\tau_C$) the subgraphs of $\tau$ consisting of
vertices  $V_{\tau}^C$  (resp.  $V_{\tau,C})$ and incident flags.
Put
$$ 
  \Delta :\, H\to H\otimes H, \quad\Delta ([\tau ]) := \sum_C [\tau^C]\otimes [\tau_C],
$$
sum being taken over all cuts of $\tau .$

\medskip
{\bf Claim.}  {\it (i) $m$ defines on $H$ the structure of a commutative 
$k$--algebra with unit $[\emptyset ]$. Set
$
\eta :\,k\to H,\, 1_k\mapsto [\emptyset ]\, .
$
\smallskip
(ii) $\Delta$ is a  coassociative comultiplication on $H$,
with counit 
$$
\varepsilon :\,H\to k,\ \sum_{\tau \in Fl} a_{[\tau ]}[\tau] \mapsto  
a_{[\emptyset ]}
$$
(iii) $(H,m,\Delta ,\varepsilon ,\eta )$ is a commutative bialgebra
with unit and counit.}
\medskip
 {\bf 5.2.1. Theorem.} (K. Ebrahimi--Fard, D. Manchon, [E-FMan]). {\it  $H$ is a Hopf algebra
(i.~e. has a unique antipode)
if one can introduce an grading on $H$  such that
$$
m(H_p\otimes H_q) \subset H_{p+q},\quad \Delta (H_n)\subset \oplus_{p+q=n} H_p\otimes H_q,
$$
and moreover,
$H_0 =k[\emptyset ]$ is one--dimensional, so that  $H$ is connected.}

\medskip

A possible choice of such grading:
$$
H_n:=\ the\ k-submodule\ of\ H\ spanned\ by\ [\tau ]\ in\ Fl\  with\ |F_{\tau}|=n. 
$$
\bigskip
\centerline{\bf 6. Regularization and renormalization }

\medskip

{\bf 6.1. Regularization by ``minimal subtraction''.} Generally, by regularization we
mean ``producing a finite answer from infinite one''. A typical example is this.

\smallskip
Consider the ring $\A$  ring of germs of meromorphic functions of $z$ at $z=0$.
Put $\A_- :=z^{-1} \bold{C} [z^{-1}]$,  and denote by
$\A_+$ the ring of germs of regular functions at $z=0$.
The value of regular function at zero is
$\varepsilon_{\A} (f):= f(0)$. Any germ is unique sum of regular one and one belonging to  $\A_- $.
\smallskip

If a function is not necessarily regular, the regularized value of $f$ at $0$ is $\varepsilon_{\A} (f_+)=f_+(0)$ where
$$
f_+(z):= f(z) - \{the\ polar\  part\ of\  f\}.
$$

Generally, a ``minimal subtraction algebra'' is a commutative associative $K$--algebra $\Cal{A}$ represented as the direct sum of
two linear subspaces  $\Cal{A}=\Cal{A}_+\oplus  \Cal{A}_-$, each being a subalgebra.
Usually $\Cal{A}$ is unital and    $1\in\Cal{A}$;  besides,  we have an augmentation homomorphism
$\varepsilon_{\Cal{A}}:\,  \Cal{A}_+\to K$.
\medskip

{\bf 6.2. Connes--Kreimer renormalization.} This is a version of regularization that:

\smallskip

(i) is performed simultaneously for an infinite family of functions indexeded by flowcharts;
\smallskip
(ii)  uses  the ``division by the collective pole part'' in a noncommutative group in place of
subtraction of an individual pole.

\smallskip
More precisely, consider   
a Hopf $K$--algebra $\H$, and
   a minimal subtraction unital algebra $\A_+,\A_-\subset \A$, $\varepsilon_{\A}:\A\to K.$ 
  
 \smallskip
  
Denote by  $G(\A )$ the group of $K$--linear maps
$\varphi :\,\H\to\A$ such that $\varphi (1_{\H})=1_{\A}$,
 with the convolution product
$$
\varphi*\psi (x):= m_{\A}(\varphi\otimes \psi )\D (x)= 
\varphi(x) +
\psi (x)+ \sum_{(x)}\varphi (x^{\prime})\psi (x^{\prime\prime})'
$$
identity $e(x):= u_{\A}\circ \varepsilon (x)$, and inversion
$$
\varphi^{*-1}(x)= e(x)+\sum_{m=1}^{\infty} (e-\varphi )^{*m}(x)
$$
\smallskip
In  situations that we will consider,  for any $x\in\r{ker}\,\varepsilon$  the latter sum contains only finitely many
non--zero summands.

\smallskip

We will say that $\varphi$ is {\it a character} if it is a homomorphism of algebras.
\smallskip

Following  Birkhoff, we  may define now ``collective pole'' and ``collective regular part'' of $\varphi$.
More precisely,
 if $\A$
is a minimal subtraction algebra,  each $\varphi\in G(\A )$
admits a unique decomposition of the form
$$
\varphi =\varphi_-^{*-1}*\varphi_+;\quad \varphi_- (1)=1_{\A},\ \varphi_{-} (\r{ker}\,\varepsilon )
\subset \A_-,\ \varphi_+(\H)\subset \A_+.
$$
Values of renormalized polar (resp. regular) parts $\varphi_{-}$
(resp. $\varphi_+$) on $\r{ker}\,\varepsilon$ are given by the inductive formulas
$$
\varphi_-(x)=-\pi \left(\varphi(x) +
 \sum_{(x)}\varphi_- (x^{\prime})\varphi (x^{\prime\prime})\right),
$$
$$
\varphi_+(x)=(\r{id}-\pi ) \left(\varphi(x) +
 \sum_{(x)}\varphi_- (x^{\prime})\varphi (x^{\prime\prime})\right).
$$

\smallskip
Here $\pi :\,\A\to \A_-$ is the polar part projection in
the algebra $\A$.

\smallskip

Physicists invented these inductive formulas: they are known as BPZH--renormalization,
for Bogolyubov--Parasyuk--Zimmermann--Hepp.

\medskip

{\bf 6.3. Deforming the Halting Problem.} Let $f$ be a partial recursive function.The Halting Problem
for $f$  is that of
recognizing whether
a number $k\in \Z$ belongs to its definition
domain $D(f)$. In this subsection, we will translate it into
 the problem, whether an analytic function $\Phi (k,f; z)$
of a complex parameter $z$  has a pole at $z=1$.
\smallskip

The relevant  minimal subtraction algebra will be a version of our example from 6.1.
\smallskip
Let $\Cal{A}_+$ be the algebra of analytic functions
in $|z|<1$, continuous at $|z|=1$,
$\varepsilon_{\Cal{A}}:\, \Phi (z)\mapsto \Phi (1).$ 
Put
$\Cal{A}_-:= (1-z)^{-1}\C [(1-z)^{-1}],$ 
$\Cal{A}:=\Cal{A}_+  \oplus \Cal{A}_-$.

\smallskip
We now choose an appropriate programming method $P$ and construct its Hopf algebra.
Basically, $\H=\H_P$ is the symmetric algebra, spanned by
isomorphism classes $[p]$ of certain descriptions. Comultiplication in $\H_P$
is dual to the composition of descriptions.

\smallskip

The main choice  is that of characters, corresponding to the halting problem.

\smallskip
 The character
$\varphi_k :\, \H_P\to \Cal{A}$ corresponding to the halting
problem at a point $k\in \Z$ for the  partial recursive
function computable with the help of a description
$p\in P(\Z,\Z)$, will be defined as 
$
\varphi_k ([p]):= \Phi (k,f;z)\in \Cal{A}
$
where the function $\Phi$ is described below.

\smallskip
Using the trick used in the theory of quantum computation
(usually applied in the context of finite automata)
we will first reduce
the general halting problem to
the recognition of fixed points of permutations.

\smallskip
Start with  a partial recursive function $f:\,X\to X$, where $X$ is a constructive world. Extend $X$ by
one point, i.~e. form $X\coprod \{*_X\}$. Choose a total recursive structure
of an additive group without torsion on 
$X\coprod \{*_X\}$ with zero $*_X$.
Extend  ${f}$ to the everywhere defined
function  $g:\,X\coprod \{*_X\}\to X\coprod \{*_X\}$, by
$
g(y):= *_X\ \r{if}\ y\notin D({f}).
$
Define 
$$
\tau_f:\,(X\coprod \{*_X\})^2 \to (X\coprod \{*_X\})^2,\quad
\tau_f(x,y):= (x+g(y),y).
$$
It is a permutation. Since $(X\coprod\{*_X\}, +)$
has no torsion, the only finite orbits
of $\tau_f^{\bold{Z}}$ are fixed points. 
\smallskip
Moreover, the restriction of
$\tau_f$ upon the recursive enumerable subset
$
D(\sigma_f):= (X\coprod \{*_X\})\times D(f)
$
induces a partial recursive permutation $\sigma_f$ 
of this subset. 
Since $g(y)$ never takes the zero value $*_X$ on
$y\in D(f)$, but always is zero outside it,
the complement to $D(\sigma_f)$ in $Y$ consists
entirely of fixed points of $\tau_f$. 

\smallskip
Thus, the halting problem for $f$
reduces to the  fixed point recognition for $\tau_f$. 

\medskip

{\bf 6.4.  The Halting Problem renormalization character.}
Define a Kolmogorov numbering on a constructive world
$X$ as a bijection  $\bold{K}=\bold{K}_u:\, X\to \Z$ arranging elements of $X$
in the increasing order of their complexities $K_u$.

\smallskip
Let  $\sigma :\,X\to X$ be a partial recursive map,
such that $\sigma$ maps $D(\sigma )$ to $D(\sigma )$
and induces a permutation of this set. Put
$
\sigma_{\bold{K}} := \bold{K}\circ \sigma\circ \bold{K}^{-1}
$
and consider this as a permutation of  the subset
$$
D(\sigma_{\bold{K}}):=\bold{K}(D(\sigma )) \subset \Z
$$
consisting of numbers of elements of $D(\sigma )$ in 
the Kolmogorov order. 

\smallskip
If  $x\in D(\sigma)$ and if the orbit $\sigma^{\bold{Z}}(x)$
is infinite, then there exist such constants $c_1,c_2 >0$ that
for $k:=\bold{K}(x)$ and all $n\in \bold{Z}$ we have
$$
c_1\cdot \bold{K}(n)\le \sigma_{\bold{K}}^n(k)\le c_2\cdot \bold{K}(n).
$$
\smallskip
Now let
 $X=\Z$ and let $\sigma$ be a partial recursive map, inducing a permutation on its definition
domain. Put
$$
\Phi (k, \sigma ;z):=\frac{1}{k^2}+\sum_{n=1}^{\infty} \frac{z^{\bold{K}(n)}}{(\sigma_{\bold{K}}^n(k))^2}.
$$
Then we have:
\medskip
{\bf 6.4.1. Theorem.} {\it (i) If $\sigma$--orbit of $x$ is finite, then
$\Phi(x, \sigma ;z)$ is a rational function in $z$ whose all poles are of the first order and lie at roots of unity.

\smallskip
(ii) If this orbit is infinite,
then $\Phi (x,\sigma ;z)$ is the Taylor series of a function analytic at $|z|<1$
and continuous at the boundary $|z|=1$.}

\bigskip

\centerline{\bf REFERENCES}

\medskip

[BoMan] D.~Borisov, Yu.~Manin.  {\it Generalized operads and their inner cohomomorhisms.}
 In: Geometry and Dynamics of Groups
and spaces (In memory of Aleksander Reznikov). Ed. by M. Kapranov et al.
Progress in Math., vol. 265. 
Birkh\"auser, Boston, pp. 247--308.
Preprint math.CT/0609748

\smallskip
[CaSt] Ch.~S.~Calude, L.~Staiger. {\it  On universal computably enumerable prefix codes.}
 Math. Struct. in Comput. Sci. 19 (2009), no. 1, 45--57. 
\smallskip

[ConKr] A.~Connes, D.~Kreimer. {\it Renormalization in quantum field  theory
and the Riemann--Hilbert problem. I. The Hopf algebra structure of
graphs and the main theorem.} Comm. Math. Phys. 210, no. 1 (2000),
249--273.

\smallskip

[De]  S.~Dehaene. {\it The Number Sense. How the Mind creates Mathematics.}
Oxford UP, 1997.

\smallskip

[DeMe] S.~Dehaene, J.~Mehler. {\it Cross--linguistic regularities
in the frequency of number words.} Cognition, 43 (1992), 1--29.

\smallskip

[Del] J.--P.~Delahaye. {\it Les entiers ne naissent pas \'egaux.}
Pour la Science, no.~421, Nov.~2012, 80--85.

\smallskip

[E-FMan] K.~Ebrahimi--Fard and D.~Manchon. {\it The combinatorics of
Bogolyubov's recursion in renormalization.} math-ph/0710.3675

\smallskip

[Lev1] L.~A.~Levin, {\it Various measures of complexity for finite objects (axiomatic
description)}, Soviet Math. Dokl. Vol.17 (1976) N. 2, 522--526.
\smallskip
[Lev2] L.~A.~Levin, {\it Randomness conservation inequalities; information
and independence in mathematical theories}, Information and Control, Vol. 61 (1984)
15--37.

\smallskip

[LiVi] Ming Li, P.~Vit\'anyi. {\it An introduction to Kolmogorov complexity
and its applications.} Springer, 1993.

\smallskip

[Mand]  B.~Mandelbrot. {\it An information theory of the statistical
structure of languages.} In Communication Theory (ed. by W.~Jackson, 
pp. 486--502, Butterworth, Woburn, MA, 1953.

\smallskip
[Ma1] D.~Yu.~Manin. {\it Zipf's Law and Avoidance of Excessive Synonymy.} Cognitive
Science, vol.~32, issue 7 (2008), pp. 1075--1078. arXiv:0710.0105.

\smallskip

[Ma2]  D.~Yu.~Manin. {\it Mandelbrot's model for Zipf's Law. Can  Mandelbrot's model explain Zipf's Law
for language?} Journ. of Quantitative Linguistics, vol.16, No. 3 (2009), 274--285.

\smallskip

[Man1] Yu.~I.~Manin. {\it A Course in Mathematical Logic for Mathematicians.} Second Edition.
Graduate  Texts in Mathematics, Springer Verlag, 2010.

\smallskip

[Man2]  Yu.~Manin. {\it
A computability challenge: asymptotic bounds and isolated
 error-correcting codes.} In: WTCS 2012 (Calude Festschrift), Ed. by M.J. Dinneen et al.,
 LNCS 7160,  pp. 174Ð182, 2012. Preprint arXiv:1107.4246
\smallskip

[Man3]  Yu.~Manin. {\it Classical computing, quantum computing,
and Shor's factoring algorithm.}  S\'eminaire Bourbaki, no. 862 (June 1999),
Ast\'erisque, vol 266, 2000, 375--404.
 quant-ph/9903008.
\smallskip
 [Man4] Yu.~Manin. {\it Renormalization and computation I. Motivation and background.}
In: Proceedings OPERADS 2009, eds. J. Loday and B. Vallette,
S\'eminaires et Congr\`es 26, Soc. Math. de France, 2012, pp. 181--223.
math.QA/0904.492
\smallskip

[Man5] Yu.~Manin. {\it Renormalization and computation II:
Time cut--off and the Halting Problem.} In: Math. Struct. in Comp. Science, vol.~22, 
Special issue, pp. 729--751, 2012, Cambridge UP. math.QA/0908.3430

\smallskip

[Man6] Yu.~Manin. {\it Kolmogorov complexity
as a hidden factor of  scientific discourse:
from Newton's law to data mining.} 
Talk at the Plenary Session of the Pontifical Academy of Sciences on
``Complexity and Analogy in Science: Theoretical, Methodological and Epistemological Aspects'', Vatican, 
November 5--7, 2012. arXiv:1301.0081

\smallskip
[Man7] Yu.~Manin. {\it Zipf's law and L.~Levin's probability distributions.}  Preprint arXiv:1301.0427

\smallskip

[ManMar] Yu.~Manin, M.~Marcolli. {\it Kolmogorov complexity and the asymptotic bound for error-correcting
codes}. Preprint arXiv:1203.0653

\smallskip

[MurSo] B.~C.~Murtra, R.~Sol\'e. {\it On the Universality of Zipf's Law.}  (2010),
Santa Fe Institute.{\it  (available online).}

\smallskip

[Sc] D.~Scott. {\it The lattice of flow diagrams.} In: Symposium on Semantics of
Algorithmic Languages, Springer LN of Mathematics, 188 (1971), 311--372.

\smallskip
[Ta] T.~Tao. {\it E pluribus unum: From Complexity, Universality.} Daedalus, Journ. of the AAAS, Summer 2012, 
23--34.

\smallskip

[Ve] Todd L.~Veldhuizen. {\it Software Libraries and Their Reuse: Entropy, Kolmogorov Complexity,
and Zipf's Law.} arXiv:cs/0508023

\smallskip

[VlaNoTsfa] S.~G.~Vladut, D.~Yu.~Nogin, M.~A.~Tsfasman. {\it  Algebraic geometric codes: basic notions.} Mathematical Surveys and Monographs, 139. American Mathematical Society, Providence, RI, 2007.

\smallskip

[Ya] N.~S.~Yanofsky. {\it Towards a definition of an algorithm.}   J.~Logic Comput. ~21 (2011), no. 2, 253--286.
math.LO/0602053

\smallskip

[Zi1] G.~K.~Zipf. {\it The psycho--biology of language.} London, Routledge, 1936.

\smallskip
[Zi2] G.~K.~Zipf. {\it Human behavior and the principle  of least effort.} Addison--Wesley, 1949.
\smallskip

[ZvLe]  A.~K.~ Zvonkin,  L.~A.~ Levin. {\it
The complexity of finite objects and the basing of the concepts of information and randomness on the theory of algorithms. }(Russian)
Uspehi Mat. Nauk 25,  no. 6(156) (1970), 8--127.

\enddocument